\documentclass[fleqn,twoside]{article}

\usepackage[headings]{espcrc2}

\usepackage{amsmath,amssymb}
\usepackage{nicefrac}
\usepackage{comment}

\usepackage{definitions}

\title{Axion-dilaton cosmology, Ricci flows and integrable structures}

\author{Domenico Orlando \address[]{Universit\`a di Milano-Bicocca and INFN,  Sezione di Milano-Bicocca,  \\
    P.zza della Scienza, 3; \\ I-20126 Milano, Italy}%
  \thanks{The author is supported in part by INFN and MIUR under
    contract 2005-024045 and by the European Community's Human
    Potential Program MRTN-CT-2004-005104.}}

\runtitle{Axion-dilaton cosmology, Ricci flows and integrable structures}
\runauthor{D. Orlando}

\begin{document}

\begin{abstract}
  In this work, based on~\cite{Bakas:2006bz}, we study
  renormalization-group flows by deforming a class of conformal
  sigma-models. At leading order in $\alpha'$, renormalization-group
  equations represent a Ricci flow. In the three-sphere
  background, the latter is described by the Halphen system, which is
  exactly solvable in terms of modular forms. The round sphere is
  found to be the unique perturbative infra-red fixed point at one
  loop order.% \vspace{1pc}
\end{abstract}

\maketitle

Conformal invariance is one of the most powerful tools in string
theory, allowing to go beyond the usual low-energy limits. On the
other hand, it imposes stringent constraints on the theory. A concrete
example is given by Wess--Zumino-Witten (\textsc{wzw}) models on
compact groups, where the presence of the topological Wess--Zumino
term gives rise to a Dirac monopole quantization (the underlying
\textsc{cft} is rational). The radius is therefore quantized and this
creates an obstruction when trying to let it vary continuously. A
possible way out would be to abandon conformal invariance. The
perturbed sigma-model is no longer an infra-red fixed point and
various parameters (such as the radius in the above example) run with
the renormalization-group (\textsc{rg}) energy scale.  In this work,
we study the behaviour of a class of three dimensional systems living
in a neighbourhood of the moduli space of the $SU(2)$ \textsc{wzw}
model. In particular we find that their one-loop dynamics is described
by a Halphen system that can be explicitly solved in terms of modular
functions.

\bigskip

Let us consider a \textsc{wzw} model on a generic compact semi-simple
group manifold $G$ at level $k$. The standard Killing form is $ \di s^2 = k
\delta_{\alpha\beta} J^\alpha \otimes J^\beta $, where $J^\alpha $ are
the usual left currents. We will consider a deformed \textsc{wzw}
model with metric
\begin{equation}
  \di s^2 = g_{\alpha\beta} J^\alpha \otimes J^\beta = k \gamma_\alpha (\mu) \delta_{\alpha \beta } J^\alpha \otimes J^\beta \, ,
\end{equation}
where $\gamma_\alpha (\mu)$ are arbitrary positive functions and $\mu$
is the \textsc{rg}-scale. The $B$ field remains unperturbed, being a
topological term.

In a dimensional-regularization scheme, the one-loop \textsc{rg}-flow
equations read~\cite{Braaten:1985is,Friedan:1980jm,Osborn:1989bu}:
\begin{equation}
  \frac{\di g_{\alpha \beta }}{\di \log \mu} = \frac{1}{2\pi}\left(R_{\alpha \beta } -
  \frac{1}{4} H_{\alpha
  \beta}^2\right) =  \frac{1}{2 \pi}{R^{-}}_{\alpha \beta } \, ,
\end{equation}
where $R_{\alpha\beta}$ are the components of the Ricci tensor, $H =
\di B$ and $H_{\alpha \beta}^2 = H_{\alpha \gamma \delta}
H_{\beta}^{\phantom{\beta} \gamma \delta}$.  We observe that the
\textsc{rg}-flow equations for the perturbation pattern at hand are
governed by a Ricci flow with the connection
\begin{equation}
  {\Gamma^-}\ud{\mu}{\nu \rho} = \Chris{\mu}{\nu \rho} - \frac{1}{2} H\ud{\mu}{\nu \rho} \, .
\end{equation}

Although the full analysis is tractable for any compact group $G$, we
will here focus on the $SU(2)$, where the flow equations can be solved
explicitly. The metric has now only three entries: $\gamma_1 (\mu),
\gamma_2 (\mu), \gamma_3 (\mu)$.
\begin{comment}
\begin{equation}
  g_{\alpha \beta} = k
  \begin{pmatrix}
    \gamma_1 (\mu) \\
    & \gamma_2 (\mu) \\
    & & \gamma_3 (\mu)
  \end{pmatrix}. \label{eq:defmets3}
\end{equation}
\end{comment}
In the vielbein of the currents we obtain a diagonal Ricci tensor with
entries
\begin{align}
  R_{11}=  \frac{\gamma_1^2 - \left( \gamma_2 - \gamma_3 \right)^2}{2 \gamma_2
      \gamma_3} \, , && \text{and permutations,}
\end{align}
and similarly for the Kalb--Ramond term which is
diagonal and reads $ \left(H^2_{11}\right)= 2 / \left( \gamma_2
  \gamma_3 \right) $ and permutations.  Introduce an
\textsc{rg}-time pointing towards the infra-red,
\begin{equation}
    \di \tilde t = - \frac{1}{2\pi \gamma_1(\mu) \gamma_2 (\mu) \gamma_3(\mu)} \di \log \mu \, ,
  \end{equation}
where we have also reabsorbed the product of the three
$\gamma_\alpha$'s. Putting everything
together one obtains the following \textsc{rg} equations:
\begin{equation}
  \label{eq:gamtor}
  2 \dfrac{1}{\gamma_1} \dfrac{\di \gamma_1}{\di \bar t}= \left(\gamma_2 - \gamma_3
  \right)^2 - \gamma_1^2 + 1 \, , \quad \text{and perm.}
\end{equation}

In the absence of torsion, the last constant term in (\ref{eq:gamtor})
is missing and the flow converges towards a round sphere of
\emph{vanishing} radius. The presence of torsion does not alter
this behaviour but affects the radius of the sphere which stabilizes
to $\sqrt{k}$ because all $\gamma_\alpha$'s now converge to one. This
non-trivial infra-red fixed point corresponds to the $SU(2)_k$
\textsc{wzw} model. Such results on the convergence of the flow are
based on asymptotic analysis. However the Ricci-flow equations
(\ref{eq:gamtor}) can be solved explicitly in the case at
hand. Indeed, setting $\tilde{t} = \log (T + T_0) $, and
\begin{equation}
  \Omega_1 = \frac{\gamma_2 \gamma_3}{T + T_0} \, , \, \Omega_2 = \frac{\gamma_3 \gamma_1}{T + T_0} \, ,\,  \Omega_3 = \frac{\gamma_1 \gamma_2}{T + T_0} \, , \label{eq:6}
\end{equation}
equations (\ref{eq:gamtor}) are recast as:
\begin{equation}
  \label{eq:halphen}
  \frac{\di \Omega_1}{\di T} = \Omega_2 \Omega_3 - \Omega_1 \left(\Omega_2
      + \Omega_3 \right) \, , \quad \text{and perm.}
\end{equation}
This is the celebrated Halphen system that was studied in the 19th
century\footnote{This observation was also made by K. Sfetsos in
  unpublished work.}. Three different scenarios are possible:

\noindent\textbf{Case 1.} The case of $\Omega_1=\Omega_2=\Omega_3$
corresponds to fully isotropic deformations. In this case, the
solution is unique: $\Omega_\alpha (T) = \nicefrac{1}{T+A}$.

\noindent\textbf{Case 2.} The most general solution with $\Omega_2 =
\Omega_3 $ is written as:
\begin{equation}
  \label{eq:polsol}
  \begin{cases}
    \Omega_1 (T) = \frac{1}{T+A}+\frac{C}{(T+A)^2}, \\
    \Omega_2 (T) = \Omega_3 (T) = \frac{1}{T+A}.
  \end{cases}
\end{equation}
This describes axisymmetric deformations of the three-sphere, namely
deformations preserving an $SU(2) \times U(1)$.

\noindent\textbf{Case 3.} When $\Omega_1 \neq \Omega_2 \neq \Omega_3 $, a
class of solutions is expressed as~\cite{Takhtajan:1992qb}:
\begin{equation}
  \label{eq:Ealpha}
  \Omega_\alpha(T) = - \frac{1}{2} \frac{\di }{\di  T } \log E_\alpha
  (iT) ,
\end{equation}
where $E_\alpha (z) $ form a triplet of modular forms of weight two
for $\Gamma(2) \subset PSL(2,\mathbb{Z})$ (see \textit{e.g.}
\cite{ford}).  We can write the $E_\alpha$ as:
\begin{equation}\label{eq:parteisenhal}
  E_1 = \frac{\nicefrac{\di \lambda}{\di z}}{\lambda} \, ,
  \, E_2 = \frac{\nicefrac{\di \lambda}{\di z}}{\lambda -1} \, ,
   \, E_3 =\frac{\nicefrac{\di \lambda}{\di z}}{\lambda(\lambda
   -1)} \, ,
\end{equation}
where $\lambda $ is the elliptic modular form $ \lambda = \vartheta_2^4 / \vartheta_3^4 $.

The modular properties of the functions under consideration set
stringent constraints between the asymptotics of the solution and
its initial conditions: large-$T$ and small-$T$ regimes are
related by $T \leftrightarrow \nicefrac{1}{T}$. Assuming
$\Omega_\alpha^0\equiv \Omega_\alpha^{\vphantom 0}(0)$
\emph{finite}, the
asymptotic behaviour ($T \to \infty$) is found to be
\emph{universally}
%\footnote{The
%Jacobi functions do not follow from positive initial values and
%consequently $\Omega_1 <0$ while $\Omega_2, \Omega_3 > 0$. This is
%not in contradiction with the existence of well behaved
%all-positive solutions.}
%\begin{equation}
%  \log E_\alpha (iT) \sim -2 \log T + \mathrm{const.\ }
%\end{equation}
%and hence
\begin{equation}\label{eq:largeT}
  \Omega_\alpha(T) \sim \frac{1}{T} + \text{subleading terms.}
  %\mathcal{O}\left(\mathrm{e}^{-T}\right).
\end{equation}
This provides an elegant proof of the universality of generic
$SU(2)$ Ricci flows in the presence of torsion towards the
corresponding \textsc{wzw} infra-red fixed point.

\bigskip 

\begin{comment}
  One can further assume $\Omega_\alpha^0$ \emph{finite and
    positive}. This is natural for describing the initial deformation
  of a three-sphere and sufficient to show that \emph{all
    $\Omega_\alpha$ remain positive} at any later time. Indeed,
  suppose that $0<\Omega_1^0<\Omega_2^0<\Omega_3^0$ and that
  $\Omega_1$ has reached at time $t_1$ the value $\Omega_1^1=0$, while
  $\Omega_2^1, \Omega_3^1>0$. From Eqs.~(\ref{eq:halphen}) we conclude
  that at time $t_1$, $\dot{\Omega}_2^1=\dot{\Omega}_3^1=-\Omega_2^1\,
  \Omega_3^1<0$ and $\dot{\Omega}_1^1=\Omega_2^1\, \Omega_3^1>0$. This
  latter inequality implies that $\Omega_1$ vanishes at $t_1$ while it
  is increasing, passing therefore from negative to positive values.
  This could only happen if $\Omega_1^0$ were negative, which
  contradicts the original assumption. However, if indeed
  $\Omega_1^0<0$ and $\Omega_2^0,\Omega_3^0>0$, there is a time $t_1$
  where $\Omega_1$ becomes positive and remains positive together with
  $\Omega_2$ and $\Omega_3$ until they reach the asymptotic region,
  where they all satisfy (\ref{eq:largeT}). The generic behaviour of a
  positive-initial-value solution is given in
  Fig.~\ref{fig:gen-sol-halph}.
  \begin{comment}
    \begin{figure}
      \begin{center}
        \includegraphics[width=.6\linewidth]{GenericHalphen}
      \end{center}
      \caption{A generic solution of the Halphen system for positive
        $\Omega_1^0<\Omega_2^0<\Omega_3^0$.}
      \label{fig:gen-sol-halph}
    \end{figure}
  \end{comment}
  Although the Halphen system can be solved, its deep nature makes it
  difficult to establish the correspondence between a given set of
  initial conditions and the modular forms $E_\alpha$ (see Eqs.
  (\ref{eq:Ealpha}) or (\ref{eq:omreal})) necessary to provide the
  actual solution.  Moreover, it is important to stress that solutions
  exist, for which the initial conditions $\Omega_\alpha^0$ are not
  all positive and not even finite. Hence, for these solutions,
  (\ref{eq:largeT}) does not hold. This happens \emph{e.g.} for the
  particular solution given in Eqs.  (\ref{eq:parteisenhal}) and
  (\ref{lamell}), where $T=0$ is a simple pole with positive (unit)
  residue for $\Omega_2, \Omega_3$ and a double pole with negative
  residue ($-\nicefrac{\pi}{2}$) for $\Omega_1$. As a consequence,
  $\Omega_1$ is negative and increases, at large $T$, exponentially
  towards zero, while $\Omega_3$ is positive and decreases
  exponentially towards zero; $\Omega_2$ is positive and decreases
  exponentially towards $\nicefrac{\pi}{2}$. Solutions with negative
  $\Omega$'s were considered in
  Refs. \cite{Atiyah:1985dv,Gibbons:1986df} for the description of the
  configuration manifold of two $SU(2)$ monopoles.

  The existence of poles is generic for \emph{all} solutions of the
  Halphen system. For solutions corresponding to a set of finite and
  different initial values, these poles are pushed behind $T=0$.  This
  is related to the following general property: Halphen's solutions
  possess a natural movable boundary \cite{Takhtajan:1992qb}. They
  exist in a domain of $\mathbb{C}$, where they are holomorphic and
  single valued, and this domain has a boundary that contains a dense
  set of essential singularities.  The precise location of this
  boundary depends on the initial conditions.

  The constants $(A,C)$ are arbitrary and determined by the initial
  conditions. This class is closed under $PSL(2, \setC )$
  transformations. Indeed, using (\ref{eq:modtr}), we learn that under
  $T\to \nicefrac{1}{T}$, $(A,C)\to
  (\nicefrac{1}{A},\nicefrac{-C}{A^2})$, while for $T\to T+1$,
  $(A,C)\to(A+1,C)$. Regularity requires $\Omega_{2,3}^0\equiv
  \nicefrac{1}{A}>0$, which ensures that the pole of $\Omega_\alpha$
  is located at $T<0$. Furthermore, the asymptotic behaviour is again
  universal as in Eq. (\ref{eq:largeT}). No new fixed point appears
  therefore in this case either. The \textsc{rg} flow of perturbations
  of this type, namely with $\gamma_1 \neq \gamma_2= \gamma_3$, was
  analyzed in \cite{Fateev:1996ea}, without torsion though. When
  translated into the language of self-dual four-dimensional Euclidean
  metrics with $SU(2)$ isometry, solutions (\ref{eq:polsol})
  correspond to the general Taub--NUT family, including Eguchi--Hanson
  metrics \cite{Gibbons:1979xn}.
\end{comment}
\begin{comment}

To summarize and conclude the present analysis, solutions of the
Halphen system fall in three classes:
$\Omega_1\neq\Omega_2\neq\Omega_3$,
$\Omega_1\neq\Omega_2=\Omega_3$ and $\Omega_1=\Omega_2=\Omega_3$.
The corresponding three-manifolds, target spaces of the
sigma-model, are homogeneous with isometry group $SU(2)$, $SU(2)
\times U(1)$ and $SU(2)\times SU(2)$, respectively. This captures
all possible isometry groups for a general three-sphere, the
latter, most symmetric case corresponding to the usual round
sphere.
\end{comment}
The Ricci flow describing the renormalization of the
sigma-model leads unavoidably to the round sphere, which is
therefore the unique perturbative infra-red fixed point found at
one loop.

\bibliography{Biblia}

\end{document}